\documentclass[a4paper,floatfix,groupedaddress,prl,showpacs,pdftex,twocolumn,twoside]{revtex4}

\usepackage{graphicx}
\usepackage{times}
\usepackage{hypernat, hyperref}

\newlength{\figwidth}
\newcommand{\degree}{\ensuremath{^\circ}}%
\newcommand{\ie}{i.\,e.}%

\begin{document}

\title{Alternating gradient focusing and deceleration of large molecules}%

\author{Kirstin Wohlfart}%
\author{Fabian Gr{\"a}tz}%
\author{Frank Filsinger}%
\author{Henrik Haak}%
\author{Gerard Meijer}%
\author{Jochen K{\"u}pper}%
\email[Author to whom correspondence should be addressed. Email:~]{jochen@fhi-berlin.mpg.de}%
\affiliation{Fritz-Haber-Institut der Max-Planck-Gesellschaft, Faradayweg 4--6, 14195 Berlin,
   Germany}%
\date{\today}%
\begin{abstract}%
   \noindent%
   We have focused and decelerated benzonitrile (C$_7$H$_5$N) molecules from a molecular beam, using
   an array of time-varying inhomogeneous electric fields in alternating gradient configuration.
   Benzonitrile is prototypical for large asymmetric top molecules that exhibit rich rotational
   structure and a high density of states. At the rotational temperature of 3.5~K in the pulsed
   molecular beam, many rotational states are populated. Benzonitrile molecules in their absolute
   ground state are decelerated from 320~m/s to 289~m/s, and similar changes in velocity are
   obtained for excited rotational states. All measurements agree well with the outcome of
   trajectory calculations. These experiments demonstrate that such large polyatomic molecules are
   amenable to the powerful method of Stark deceleration.
\end{abstract}
\maketitle%

\noindent%
The production of cold molecular samples offers fascinating prospects for fundamental physics
studies~\cite{Daussy:PRL83:1554, Hudson:PRL89:023003} and ultracold
chemistry~\cite{Krems:IRPC24:99}. One of the many techniques developed for the production of cold
molecules~\cite{EPJD31:ColdMol} relies on the deceleration of polar molecules from a molecular beam
using an array of time-varying inhomogeneous electric fields~\cite{Bethlem:PRL83:1558}. Over the
last years, tremendous advances have been made in the Stark deceleration and trapping of small,
polar molecules in low-field-seeking (lfs) states~\cite{Bethlem:Nature406:491,
   Meerakker:PRL94:023004, Heiner:NatPhys3:115, Sawyer:PRL98:253002}. Molecules in lfs states are
confined by a minimum of the electric field, which can readily be created on the molecular beam
axis. For large molecules with a dense manifold of rotational states, however, all states are
high-field seeking (hfs) at the relevant electric field strengths. To confine such molecules, a
maximum of the electric field on the molecular beam axis would be needed. Since such a maximum
cannot be created using static electric fields, dynamic focusing -- also referred to as alternating
gradient (AG) focusing -- has to be used. In a single AG focusing lens the molecules experience a
focusing force in one transverse direction and a defocusing force in the perpendicular direction. In
an array of AG lenses that alternately focus and defocus in the transverse directions the molecules
can be dynamically focused and transported along the molecular beam axis~\cite{Auerbach:JCP45:2160}.
In order to decelerate the molecules, the longitudinal electric field gradients at the entrance and
exit of the AG lens can be exploited. In practice, for a molecules in hfs states, this is done by
switching the high voltage on while the molecules are already inside the AG lens, and by keeping the
voltages on when the molecules exit the lens. So far, only a few proof-of-principle experiments on
the AG focusing and deceleration of diatomic molecules (metastable CO and YbF) in hfs states have
been performed~\cite{Bethlem:PRL88:133003, Tarbutt:PRL92:173002, Bethlem:JPB39:R263}. Complementary,
slow beams of thermally stable large molecules can directly emerge from an
oven~\cite{Deachapunya:EPJD46:307}. However, these samples are internally hot ($\sim600$~K) and,
therefore, the population is distributed over a large number of quantum states and different
conformers.

It is important to experimentally demonstrate that AG focusing and deceleration also work for large
molecules, in order to produce cold, slow samples thereof. This would vastly extend the range of
species whose motion and orientation can be precisely controlled. In particular, this would enable
novel studies on the ``molecular building blocks of life''. During the last decades, structural and
dynamical properties of these bio-molecules have been unraveled via detailed spectroscopic studies
on isolated systems in the gas phase~\cite{EPJD20:Biomolecules, PCCP6:Biomolecules,
   Vries:ARPC58:585}. In these studies, the advances in the preparation of beams of internally cold
bio-molecules have been instrumental. The control over the velocity of the molecules provided by the
AG decelerator would enable increased observation times, and thereby, in principle, a higher
spectral resolution. This would have a possible application in the search for parity violation in
chiral molecules~\cite{Daussy:PRL83:1554}. Furthermore, the rotational state selection intrinsic to
the AG focusing and deceleration process would result in less congested spectra. In a molecular
beam, bio-molecules occur in multiple conformational structures~\cite{Suenram:JACS102:7180,
   Rizzo:JCP83:4819}. For tryptophan, for example, six conformers have been observed
\cite{Rizzo:JCP83:4819, Snoek:PCCP3:1819} with their calculated dipole moments ranging from 1~D to
7~D. The state selection in the AG decelerator would naturally provide packets of selected
conformers. These state- and conformer-selected packets of molecules can be strongly aligned or
oriented. They would, for instance, be highly beneficial for electron or X-ray diffraction imaging
studies~\cite{Hedberg:Science254:410, Chapman:NatPhys2:839}, where disentangling the overlaying
contributions from different species would be an onerous task. The tunable velocity and the narrow
velocity distribution of the AG decelerated beams would also be of great advantage for matter wave
interferometry experiments~\cite{Gerlich:NatPhys3:711}.

Here we demonstrate the AG focusing and deceleration of benzonitrile (C$_7$H$_5$N), a prototypical
large molecule. The electronic ground states of such polyatomic organic molecules are typically
closed shell (singlet) states and the molecules are asymmetric rotors, resulting in a complex and
dense rotational level structure. The Stark effect of these molecules is due to the coupling of
closely spaced rotational states by an electric field~\cite{Gordy:MWMolSpec}. Experiments are
performed with benzonitrile in various rotational states, including its absolute ground state
$J_{K_aK_c}=0_{00}$. This is the only state that is stable against collisions at sufficiently low
temperatures. Excited rotational states have the added complication of real and avoided crossings of
levels as a function of the electric field strength. It is not \emph{a priori} clear whether one can
focus and decelerate molecules in states within such a dense Stark
manifold~\cite{Vliegen:PRL92:033005, AbdElRahim:JPCA109:8507}. However, we successfully demonstrate
this for the $4_{04}$ state. All measurements agree well with the outcome of trajectory
calculations.

\begin{figure}%
   \centering%
   \includegraphics[width=\figwidth]{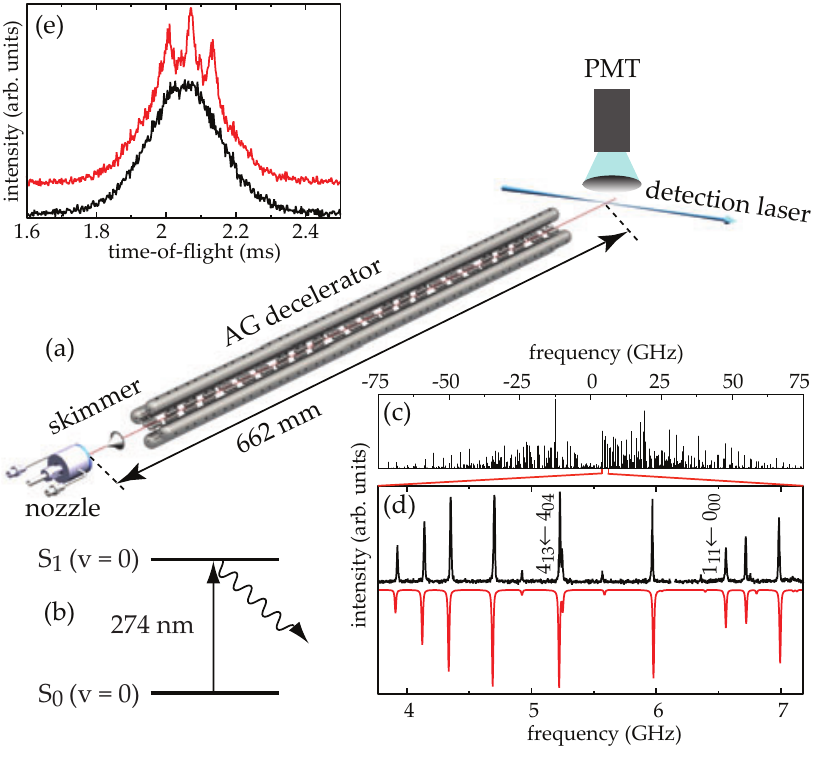}%
   \caption{(Color online) (a)~Schematic view of the experimental setup. (b)~LIF detection scheme.
      (c)~Simulated rotationally resolved electronic excitation spectrum of benzonitrile at the
      rotational temperature of 3.5~K. The frequencies are given relative to the origin at
      36512.74~cm$^{-1}$. (d)~Expanded view of the experimental (upper black trace) and simulated
      (lower red trace) spectrum. (e)~LIF signal of benzonitrile in its ground state as a function
      of time-of-flight from the nozzle to the detector. The lower (black) trace depicts the
      free-flight signal, and the upper (red) trace, which is offset for clarity, depicts the signal
      for optimum focusing conditions.}%
   \label{fig:setup}%
\end{figure}

The experimental setup is shown in Fig.~\ref{fig:setup}\,a.
Benzonitrile is co-expanded in 0.7~bar of xenon at a temperature of 300\,K through a pulsed nozzle
with a 0.8~mm orifice and an opening time of about $150$~$\mu$s. The experiments are performed at a
repetition rate of 40~Hz. The molecular beam has a velocity of approximately 320~m/s and a full
width at half maximum (FWHM) velocity spread of about 10~\%. At a distance of 27~mm downstream from
the nozzle the molecules pass through a 1.5~mm diameter skimmer mounted directly on a gate
valve~\cite{Kuepper:RSI77:016106}, and then enter the AG decelerator. The decelerator consists of 27
AG lenses arranged along the molecular beam axis. The first lens starts 37~mm behind the tip of the
skimmer. Each lens consists of a pair of 13~mm long electrodes with a diameter of 6~mm and spherical
end caps. The two parallel electrodes within each lens have an opening of 2~mm between them.
Successive lenses are placed at center-to-center distances of 20~mm and are rotated relative to each
other by 90\degree\ after every third lens, adding up to a total length of the decelerator of
533~mm. The positioning of the electrodes is accurate to about 0.1~mm. For all pairs of electrodes,
voltages are switched between $\pm15$~kV and ground. An additional bias voltage of up to $\pm0.3$~kV
is applied for the suppression of unwanted transitions between levels at zero field. The applied
high voltages create a maximum electric field strength of 143~kV/cm on the molecular beam axis.

Benzonitrile molecules are detected 662~mm behind the nozzle by laser-induced fluorescence (LIF) at
274~nm using time-resolved photon counting. A frequency-doubled and frequency-stabilized,
narrow-linewidth, continuous ring dye laser is used for state-specific detection. A part of the
experimental $S_1\!\leftarrow\!S_0$ excitation spectrum is shown in Fig.~\ref{fig:setup}\,d, where
the individual rovibronic transitions are clearly resolved. The spectrum is
simulated~\cite{Western:pgopher} using the known molecular constants~\cite{Borst:CPL350:485,
   Wohlfart:JMolSpec247:119}. Full agreement with the measured spectrum is achieved using a
rotational temperature of 3.5~K, the known natural linewidth (FWHM) of
8~MHz~\cite{Borst:CPL350:485}, and a Gaussian contribution of 7.5~MHz which accounts for Doppler
broadening and the laser linewidth. To record the time-of-flight (TOF) distribution of the molecules
in a given state from the nozzle to the detector, the laser is set to the desired transition
frequency. TOF profiles for the $0_{00}$ state are shown in Fig.~\ref{fig:setup}\,e, both for free
flight, \ie, when no voltages are applied to the decelerator, and for optimum focusing conditions.
When the high voltages are on, a structured TOF distribution is observed containing a background
that resembles the free-flight distribution. The peak intensity of the focused molecules is enhanced
by 30~\% with respect to that of the molecules in free flight. In the remainder, we show and discuss
the difference-TOF profiles, obtained by taking the difference of the measurements with and without
electric fields.

\begin{figure} \centering
   \includegraphics[width=\figwidth]{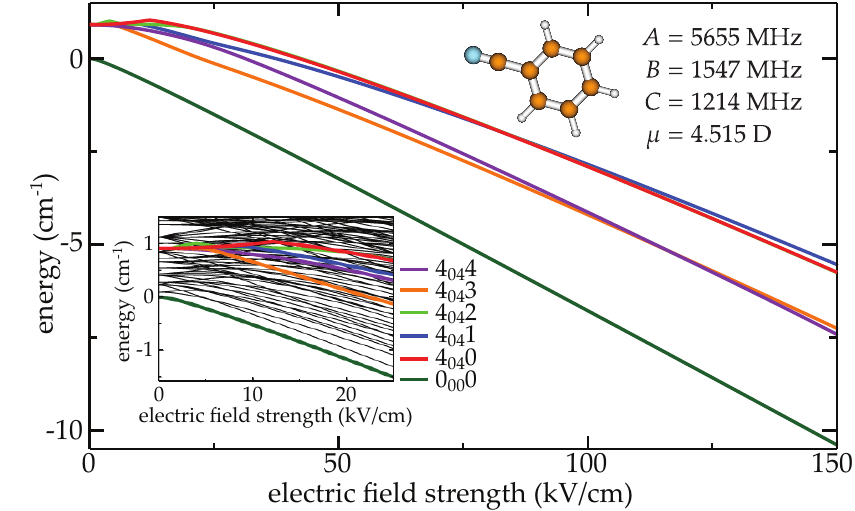}
   \caption{(Color online) Energies of all $M$ levels of the examined rotational states
      $J_{K_aK_c}M=0_{00}0$ and $4_{04}M$ ($M=0,\ldots,4$) of benzonitrile as a function of the
      electric field strength. The structure, the rotational constants, and the dipole moment of
      benzonitrile are given in the upper right corner. The inset illustrates the energies of the
      lowest levels for small electric field strengths.}
   \label{fig:stark}
\end{figure}%
The rotational constants of benzonitrile and its dipole moment are precisely
known~\cite{Wohlfart:JMolSpec247:119}. Therefore, the Stark curves of the rotational states and the
trajectories of benzonitrile in an AG decelerator can be accurately calculated. The energies of the
relevant levels are shown in Fig.~\ref{fig:stark} as a function of the electric field strength. The
rotational ground state $0_{00}$ has only a single $M$ level, whereas the $4_{04}$ state splits into
five $M$ levels. For all excited rotational states real and avoided level crossings occur at low
electric field strengths, as is shown for the $4_{04}$ state in Fig.~\ref{fig:stark}.

To describe the sequence of times at which the high voltages applied to the AG decelerator are
switched on and off, we use the concept of a synchronous molecule. By definition, the synchronous
molecule is always at the same position within an AG lens when the high voltages are switched. The
time interval during which the high voltages are applied to an AG lens, in combination with the
velocity of the synchronous molecule, determine the focusing length $f$. The amount of kinetic
energy that is removed per AG lens depends on the position of the synchronous molecule on the
molecular beam axis at the time when the high voltages are switched off. This position, relative to
the center of the AG lens, is denoted by $d$.

\begin{figure} \centering
   \includegraphics[width=\figwidth]{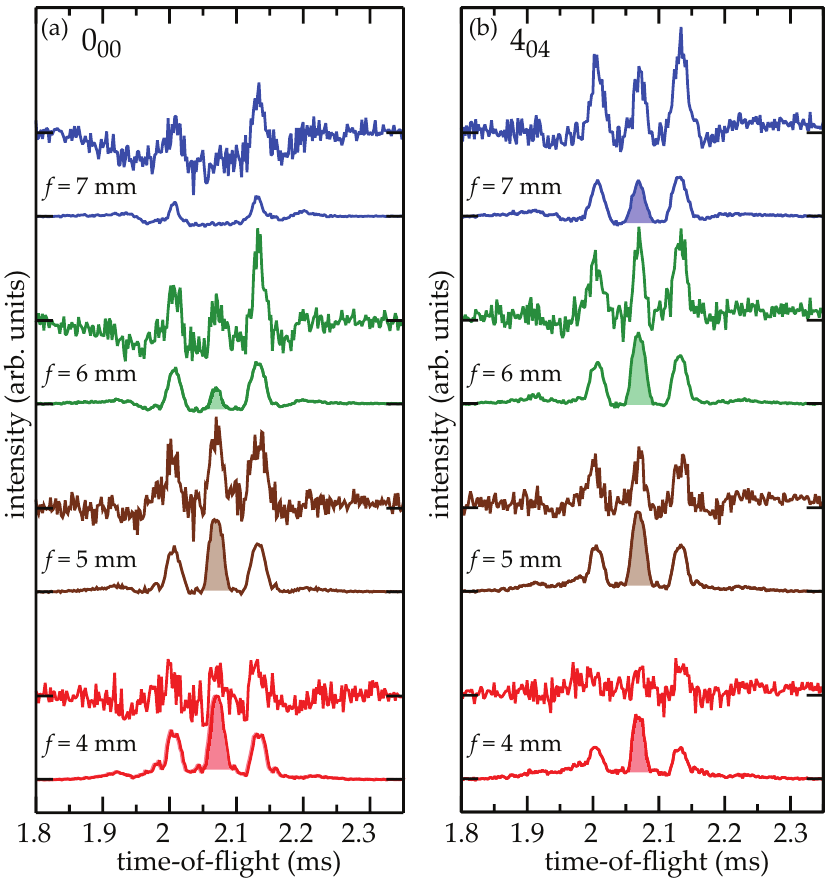}
   \caption{(Color online) Sequence of difference-TOF profiles for the (a) $0_{00}$ and (b) $4_{04}$
      states of benzonitrile obtained using different focusing lengths $f$ for a constant velocity
      of 320~m/s. The upper traces show the difference between the experimental TOF profiles with
      and without high voltages applied. The lower traces show the corresponding simulations, where
      the packets containing the synchronous molecule have been shaded.}
   \label{fig:focusing}
\end{figure}

Fig.~\ref{fig:focusing} illustrates the focusing behavior of the $0_{00}$ and $4_{04}$ states of
benzonitrile for a constant velocity of the synchronous molecule of 320~m/s.
The high voltages are switched on symmetrically around the centers of the AG lenses. Therefore, the
molecular packet is focused in both transverse directions as well as in the longitudinal direction
(bunching), but no change of the synchronous velocity occurs. In the experiments three packets of
focused molecules are observed. The central peaks of the TOF distributions occur 2.07~ms after the
molecules exit the nozzle. These packets contain the synchronous molecule. Hereafter, they are
referred to as the ``synchronous packets'', and they are shaded in the simulated TOF distributions.
The peaks at earlier and later arrival times correspond to molecular packets leading and trailing
the synchronous packet by one AG lens (or 20~mm), respectively. These packets are also focused in
all three dimensions. However, due to the lens pattern in our setup, they experience only 2/3 of the
lenses at high voltage. This results in a reduced overall focusing for these packets.

For the focusing of the $0_{00}$ state it is seen that the synchronous packet is most intense for a
focusing length of $f=5$~mm. Under these conditions approximately $10^5$ molecules per quantum state
per pulse are confined in the phase-stable central peak, corresponding to a density of
$10^8$~cm$^{-3}$. For smaller focusing lengths a shallower time-averaged confinement potential is
created, and less molecules are guided through the decelerator. For larger focusing lengths the
molecular packet is over-focused, also resulting in a decreased transmission. For $f=7$~mm the
over-focusing is so severe that the synchronous packet completely disappears. As expected, the
non-synchronous packets benefit from the increased focusing lengths. Fig.~\ref{fig:focusing}\,b
shows the focusing behavior for the $4_{04}$ state. The observed focusing effects are similar to
those for the $0_{00}$ state. Due to the smaller Stark shifts of the five $M$ levels, a larger
optimum focusing length of $f=6$~mm is found for the synchronous packet. Moreover, each of these $M$
levels has a distinct Stark curve and, therefore, a distinct focusing behavior. The measured TOFs
are the sums of these five individual contributions, resulting in a considerably weaker dependence
of the overall transmission on the exact focusing length. The simulated TOF distributions obtained
from trajectory simulations match the experimental results very well. It should be noted, that all
simulated profiles are scaled down by a factor of seventeen. Extensive simulations indicate that
this can be attributed to mechanical misalignment of the electrodes on the order of $0.1$~mm. No
additional losses for the $4_{04}$ state compared to the $0_{00}$ state are observed, demonstrating
that transitions between different states at avoided crossings are not as severe as might have been
anticipated~\cite{Vliegen:PRL92:033005, AbdElRahim:JPCA109:8507}.

\begin{figure} \centering
   \includegraphics[width=\figwidth]{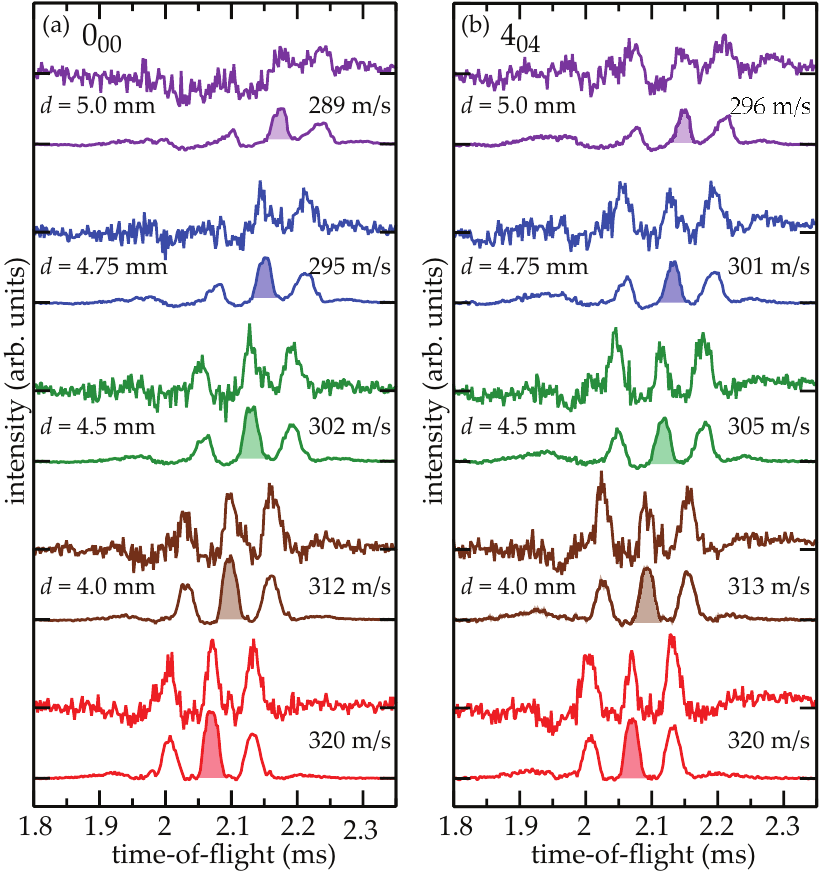}
   \caption{(Color online) Sequence of difference-TOF profiles for the deceleration of the (a)
      $0_{00}$ and (b) $4_{04}$ states of benzonitrile from an initial velocity of 320~m/s to the
      indicated final velocities. The positions $d$ that have been used to reach these final
      velocities are specified (see text for details). The upper traces show the difference between
      the experimental TOF profiles with and without high voltages applied. The lower traces show
      the corresponding simulations, where the packets containing the synchronous molecule have been
      shaded.}
   \label{fig:deceleration}
\end{figure}

Fig.~\ref{fig:deceleration} presents the results for the deceleration of benzonitrile in its
$0_{00}$ and $4_{04}$ states.
The bottommost (red) traces show focusing experiments for a constant velocity of the synchronous
molecule of 320~m/s using the optimum focusing lengths. All other traces show experiments in which
the synchronous packet is decelerated from 320~m/s to successively lower velocities, resulting in
later arrival times at the detector. For the topmost trace, recorded using $d=5$~mm, the packet is
decelerated to 289~m/s, corresponding to a reduction of the kinetic energy by 18~\%. The observed
intensities of the non-synchronous packets decrease faster upon increasing $d$. Because the
molecules in these packets miss every third deceleration stage, their trajectories are not stable
and they are only observed due to the finite length of the decelerator. When deceleration to lower
velocities is performed by increasing $d$, as is done here, the signal intensity decreases due to
the reduction in phase space acceptance. However, one could also decelerate to lower velocities by
increasing the number of AG lenses for a given value of $d$. In this case, in principle, no decrease
in the intensity of the synchronous packet is expected due to the phase stability of the
deceleration process~\cite{Bethlem:JPB39:R263}. The deceleration behavior of the $4_{04}$ state
shown in Fig.~\ref{fig:deceleration}\,b is similar to that of the ground state. However, it is
important to realize that the optimum deceleration time sequences differ for the individual $M$
levels. Using time sequences calculated for the $M=0$ level, the synchronous packet is decelerated
to successively lower final velocities, down to 296~m/s. Apart from slightly different final
velocities and intensities, analogous results are obtained for time sequences calculated for other
$M$ levels. For a considerably longer decelerator and a given time sequence, the $M$ levels could,
in principle, be separated based on their different Stark shifts. For all deceleration measurements,
the relative intensities and the arrival times of the molecular packets at the detector are nicely
reproduced by the trajectory simulations.

In conclusion, we have demonstrated that alternating gradient focusing and deceleration can be
performed for large asymmetric top molecules. We have examined the focusing properties for
benzonitrile in several low-lying rotational states. Benzonitrile has been decelerated both in its
absolute ground state and in excited rotational states. While only data for the $0_{00}$ and
$4_{04}$ states have been shown, focusing and deceleration have also been demonstrated for the
$1_{01}$ and $6_{06}$ states. If desired, different rotational states can be decelerated
simultaneously, albeit with different acceptances. It should be realized, that already in the
current experiment a significant amount of state-selection is achieved. Whereas the different
low-energy rotational states investigated in this work are focused and decelerated with similar
acceptances, higher rotational states often have negligible Stark shifts and are practically not
accepted by the decelerator.

Currently we are extending the decelerator to 81 deceleration stages, using refined mechanical
alignment procedures and an improved lens pattern, where we rotate the focusing direction every 2
stages. With this setup we expect to be able to decelerate benzonitrile molecules from 320~m/s to
150~m/s, removing about 80~\% of the kinetic energy, with intensities about an order of magnitude
smaller than the ones reported here. For the deceleration to even lower velocities and eventually
loading the molecules into a trap, hyperbolic electrode geometries will be
used~\cite{Bethlem:JPB39:R263}. Three-dimensional confinement is possible, for instance, in AC
electric traps, in which the very same principles of dynamic focusing as in the AG decelerator are
employed~\cite{Veldhoven:PRL94:083001}. This will enable us to select and decelerate conformers of
large molecules, such as the amino acid tryptophan or small model peptides, to low velocities, and
eventually trap them. This would allow to precisely study the intrinsic properties of these species,
for example, dynamical processes on timescales ranging from femto-seconds to fractions of a second,
\ie{}, interconversion between conformational structures.

\begin{acknowledgments}%
   This work would not have been possible without the technical support at the FHI. Helpful
   discussions with Hendrick L.\ Bethlem as well as financial support from the \emph{Deutsche
      Forschungsgemeinschaft} within the priority program 1116 ``Interactions in ultracold gases''
   are gratefully acknowledged.
\end{acknowledgments}

\bibliographystyle{apsrev-nourl}
\bibliography{string,mp}

\begin{thebibliography}{31}
\expandafter\ifx\csname natexlab\endcsname\relax\def\natexlab#1{#1}\fi
\expandafter\ifx\csname bibnamefont\endcsname\relax
  \def\bibnamefont#1{#1}\fi
\expandafter\ifx\csname bibfnamefont\endcsname\relax
  \def\bibfnamefont#1{#1}\fi
\expandafter\ifx\csname citenamefont\endcsname\relax
  \def\citenamefont#1{#1}\fi
\expandafter\ifx\csname url\endcsname\relax
  \def\url#1{\texttt{#1}}\fi
\expandafter\ifx\csname urlprefix\endcsname\relax\def\urlprefix{URL }\fi
\providecommand{\bibinfo}[2]{#2}
\providecommand{\eprint}[2][]{\url{#2}}

\bibitem[{\citenamefont{Daussy et~al.}(1999)\citenamefont{Daussy, Marrel,
  Amy-Klein, Nguyen, Bord\'e, and Chardonnet}}]{Daussy:PRL83:1554}
\bibinfo{author}{\bibfnamefont{C.}~\bibnamefont{Daussy}},
  \bibinfo{author}{\bibfnamefont{T.}~\bibnamefont{Marrel}},
  \bibinfo{author}{\bibfnamefont{A.}~\bibnamefont{Amy-Klein}},
  \bibinfo{author}{\bibfnamefont{C.~T.} \bibnamefont{Nguyen}},
  \bibinfo{author}{\bibfnamefont{C.~J.} \bibnamefont{Bord\'e}},
  \bibnamefont{and}
  \bibinfo{author}{\bibfnamefont{C.}~\bibnamefont{Chardonnet}},
  \bibinfo{journal}{Phys. Rev. Lett.} \textbf{\bibinfo{volume}{83}},
  \bibinfo{pages}{1554} (\bibinfo{year}{1999}).

\bibitem[{\citenamefont{Hudson et~al.}(2002)\citenamefont{Hudson, Sauer,
  Tarbutt, and Hinds}}]{Hudson:PRL89:023003}
\bibinfo{author}{\bibfnamefont{J.~J.} \bibnamefont{Hudson}},
  \bibinfo{author}{\bibfnamefont{B.~E.} \bibnamefont{Sauer}},
  \bibinfo{author}{\bibfnamefont{M.~R.} \bibnamefont{Tarbutt}},
  \bibnamefont{and} \bibinfo{author}{\bibfnamefont{E.~A.} \bibnamefont{Hinds}},
  \bibinfo{journal}{Phys. Rev. Lett.} \textbf{\bibinfo{volume}{89}},
  \bibinfo{pages}{023003} (\bibinfo{year}{2002}).

\bibitem[{\citenamefont{Krems}(2005)}]{Krems:IRPC24:99}
\bibinfo{author}{\bibfnamefont{R.~V.} \bibnamefont{Krems}},
  \bibinfo{journal}{Int. Rev. Phys. Chem.} \textbf{\bibinfo{volume}{24}},
  \bibinfo{pages}{99} (\bibinfo{year}{2005}).

\bibitem[{\citenamefont{{Special Issue ``Ultracold Polar Molecules: Formation
  and Collisions''}}(2004)}]{EPJD31:ColdMol}
\bibinfo{author}{\bibnamefont{{Special Issue ``Ultracold Polar Molecules:
  Formation and Collisions''}}}, \bibinfo{journal}{Eur. Phys. J. D}
  \textbf{\bibinfo{volume}{31}}, \bibinfo{pages}{149} (\bibinfo{year}{2004}).

\bibitem[{\citenamefont{Bethlem et~al.}(1999)\citenamefont{Bethlem, Berden, and
  Meijer}}]{Bethlem:PRL83:1558}
\bibinfo{author}{\bibfnamefont{H.~L.} \bibnamefont{Bethlem}},
  \bibinfo{author}{\bibfnamefont{G.}~\bibnamefont{Berden}}, \bibnamefont{and}
  \bibinfo{author}{\bibfnamefont{G.}~\bibnamefont{Meijer}},
  \bibinfo{journal}{Phys. Rev. Lett.} \textbf{\bibinfo{volume}{83}},
  \bibinfo{pages}{1558} (\bibinfo{year}{1999}).

\bibitem[{\citenamefont{Bethlem et~al.}(2000)\citenamefont{Bethlem, Berden,
  Crompvoets, Jongma, van Roij, and Meijer}}]{Bethlem:Nature406:491}
\bibinfo{author}{\bibfnamefont{H.~L.} \bibnamefont{Bethlem}},
  \bibinfo{author}{\bibfnamefont{G.}~\bibnamefont{Berden}},
  \bibinfo{author}{\bibfnamefont{F.~M.~H.} \bibnamefont{Crompvoets}},
  \bibinfo{author}{\bibfnamefont{R.~T.} \bibnamefont{Jongma}},
  \bibinfo{author}{\bibfnamefont{A.~J.~A.} \bibnamefont{van Roij}},
  \bibnamefont{and} \bibinfo{author}{\bibfnamefont{G.}~\bibnamefont{Meijer}},
  \bibinfo{journal}{Nature} \textbf{\bibinfo{volume}{406}},
  \bibinfo{pages}{491} (\bibinfo{year}{2000}).

\bibitem[{\citenamefont{van~de Meerakker et~al.}(2005)\citenamefont{van~de
  Meerakker, Smeets, Vanhaecke, Jongma, and Meijer}}]{Meerakker:PRL94:023004}
\bibinfo{author}{\bibfnamefont{S.~Y.~T.} \bibnamefont{van~de Meerakker}},
  \bibinfo{author}{\bibfnamefont{P.~H.~M.} \bibnamefont{Smeets}},
  \bibinfo{author}{\bibfnamefont{N.}~\bibnamefont{Vanhaecke}},
  \bibinfo{author}{\bibfnamefont{R.~T.} \bibnamefont{Jongma}},
  \bibnamefont{and} \bibinfo{author}{\bibfnamefont{G.}~\bibnamefont{Meijer}},
  \bibinfo{journal}{Phys. Rev. Lett.} \textbf{\bibinfo{volume}{94}},
  \bibinfo{pages}{023004} (\bibinfo{year}{2005}).

\bibitem[{\citenamefont{Heiner et~al.}(2007)\citenamefont{Heiner, Carty,
  Meijer, and Bethlem}}]{Heiner:NatPhys3:115}
\bibinfo{author}{\bibfnamefont{C.~E.} \bibnamefont{Heiner}},
  \bibinfo{author}{\bibfnamefont{D.}~\bibnamefont{Carty}},
  \bibinfo{author}{\bibfnamefont{G.}~\bibnamefont{Meijer}}, \bibnamefont{and}
  \bibinfo{author}{\bibfnamefont{H.~L.} \bibnamefont{Bethlem}},
  \bibinfo{journal}{Nature Phys.} \textbf{\bibinfo{volume}{3}},
  \bibinfo{pages}{115} (\bibinfo{year}{2007}).

\bibitem[{\citenamefont{Sawyer et~al.}(2007)\citenamefont{Sawyer, Lev, Hudson,
  Stuhl, Lara, Bohn, and Ye}}]{Sawyer:PRL98:253002}
\bibinfo{author}{\bibfnamefont{B.~C.} \bibnamefont{Sawyer}},
  \bibinfo{author}{\bibfnamefont{B.~L.} \bibnamefont{Lev}},
  \bibinfo{author}{\bibfnamefont{E.~R.} \bibnamefont{Hudson}},
  \bibinfo{author}{\bibfnamefont{B.~K.} \bibnamefont{Stuhl}},
  \bibinfo{author}{\bibfnamefont{M.}~\bibnamefont{Lara}},
  \bibinfo{author}{\bibfnamefont{J.~L.} \bibnamefont{Bohn}}, \bibnamefont{and}
  \bibinfo{author}{\bibfnamefont{J.}~\bibnamefont{Ye}}, \bibinfo{journal}{Phys.
  Rev. Lett.} \textbf{\bibinfo{volume}{98}}, \bibinfo{pages}{253002}
  (\bibinfo{year}{2007}).

\bibitem[{\citenamefont{Auerbach et~al.}(1966)\citenamefont{Auerbach, Bromberg,
  and Wharton}}]{Auerbach:JCP45:2160}
\bibinfo{author}{\bibfnamefont{D.}~\bibnamefont{Auerbach}},
  \bibinfo{author}{\bibfnamefont{E.~E.~A.} \bibnamefont{Bromberg}},
  \bibnamefont{and} \bibinfo{author}{\bibfnamefont{L.}~\bibnamefont{Wharton}},
  \bibinfo{journal}{J. Chem. Phys.} \textbf{\bibinfo{volume}{45}},
  \bibinfo{pages}{2160} (\bibinfo{year}{1966}).

\bibitem[{\citenamefont{Bethlem et~al.}(2002)\citenamefont{Bethlem, van Roij,
  Jongma, and Meijer}}]{Bethlem:PRL88:133003}
\bibinfo{author}{\bibfnamefont{H.~L.} \bibnamefont{Bethlem}},
  \bibinfo{author}{\bibfnamefont{A.~J.~A.} \bibnamefont{van Roij}},
  \bibinfo{author}{\bibfnamefont{R.~T.} \bibnamefont{Jongma}},
  \bibnamefont{and} \bibinfo{author}{\bibfnamefont{G.}~\bibnamefont{Meijer}},
  \bibinfo{journal}{Phys. Rev. Lett.} \textbf{\bibinfo{volume}{88}},
  \bibinfo{pages}{133003} (\bibinfo{year}{2002}).

\bibitem[{\citenamefont{Tarbutt et~al.}(2004)\citenamefont{Tarbutt, Bethlem,
  Hudson, Ryabov, Ryzhov, Sauer, Meijer, and Hinds}}]{Tarbutt:PRL92:173002}
\bibinfo{author}{\bibfnamefont{M.~R.} \bibnamefont{Tarbutt}},
  \bibinfo{author}{\bibfnamefont{H.~L.} \bibnamefont{Bethlem}},
  \bibinfo{author}{\bibfnamefont{J.~J.} \bibnamefont{Hudson}},
  \bibinfo{author}{\bibfnamefont{V.~L.} \bibnamefont{Ryabov}},
  \bibinfo{author}{\bibfnamefont{V.~A.} \bibnamefont{Ryzhov}},
  \bibinfo{author}{\bibfnamefont{B.~E.} \bibnamefont{Sauer}},
  \bibinfo{author}{\bibfnamefont{G.}~\bibnamefont{Meijer}}, \bibnamefont{and}
  \bibinfo{author}{\bibfnamefont{E.~A.} \bibnamefont{Hinds}},
  \bibinfo{journal}{Phys. Rev. Lett.} \textbf{\bibinfo{volume}{92}},
  \bibinfo{pages}{173002} (\bibinfo{year}{2004}).

\bibitem[{\citenamefont{Bethlem et~al.}(2006)\citenamefont{Bethlem, Tarbutt,
  K\"upper, Carty, Wohlfart, Hinds, and Meijer}}]{Bethlem:JPB39:R263}
\bibinfo{author}{\bibfnamefont{H.~L.} \bibnamefont{Bethlem}},
  \bibinfo{author}{\bibfnamefont{M.~R.} \bibnamefont{Tarbutt}},
  \bibinfo{author}{\bibfnamefont{J.}~\bibnamefont{K\"upper}},
  \bibinfo{author}{\bibfnamefont{D.}~\bibnamefont{Carty}},
  \bibinfo{author}{\bibfnamefont{K.}~\bibnamefont{Wohlfart}},
  \bibinfo{author}{\bibfnamefont{E.~A.} \bibnamefont{Hinds}}, \bibnamefont{and}
  \bibinfo{author}{\bibfnamefont{G.}~\bibnamefont{Meijer}},
  \bibinfo{journal}{J. Phys. B} \textbf{\bibinfo{volume}{39}},
  \bibinfo{pages}{R263} (\bibinfo{year}{2006}).

\bibitem[{\citenamefont{Deachapunya et~al.}(2008)\citenamefont{Deachapunya,
  Fagan, Major, Reiger, Ritsch, Stefanov, Ulbricht, and
  Arndt}}]{Deachapunya:EPJD46:307}
\bibinfo{author}{\bibfnamefont{S.}~\bibnamefont{Deachapunya}},
  \bibinfo{author}{\bibfnamefont{P.~J.} \bibnamefont{Fagan}},
  \bibinfo{author}{\bibfnamefont{A.~G.} \bibnamefont{Major}},
  \bibinfo{author}{\bibfnamefont{E.}~\bibnamefont{Reiger}},
  \bibinfo{author}{\bibfnamefont{H.}~\bibnamefont{Ritsch}},
  \bibinfo{author}{\bibfnamefont{A.}~\bibnamefont{Stefanov}},
  \bibinfo{author}{\bibfnamefont{H.}~\bibnamefont{Ulbricht}}, \bibnamefont{and}
  \bibinfo{author}{\bibfnamefont{M.}~\bibnamefont{Arndt}},
  \bibinfo{journal}{Eur. Phys. J. D} \textbf{\bibinfo{volume}{46}},
  \bibinfo{pages}{307} (\bibinfo{year}{2008}).

\bibitem[{\citenamefont{{Special issue ``Molecular physics of building blocks
  of life under isolated or defined conditions''}}(2002)}]{EPJD20:Biomolecules}
\bibinfo{author}{\bibnamefont{{Special issue ``Molecular physics of building
  blocks of life under isolated or defined conditions''}}},
  \bibinfo{journal}{Eur. Phys. J. D} \textbf{\bibinfo{volume}{20}},
  \bibinfo{pages}{309} (\bibinfo{year}{2002}).

\bibitem[{\citenamefont{{Special issue ``Bio-active molecules in the gas
  phase''}}(2004)}]{PCCP6:Biomolecules}
\bibinfo{author}{\bibnamefont{{Special issue ``Bio-active molecules in the gas
  phase''}}}, \bibinfo{journal}{Phys. Chem. Chem. Phys.}
  \textbf{\bibinfo{volume}{6}}, \bibinfo{pages}{2543} (\bibinfo{year}{2004}).

\bibitem[{\citenamefont{de~Vries and Hobza}(2007)}]{Vries:ARPC58:585}
\bibinfo{author}{\bibfnamefont{M.~S.} \bibnamefont{de~Vries}} \bibnamefont{and}
  \bibinfo{author}{\bibfnamefont{P.}~\bibnamefont{Hobza}},
  \bibinfo{journal}{Ann. Rev. Phys. Chem.} \textbf{\bibinfo{volume}{58}},
  \bibinfo{pages}{585} (\bibinfo{year}{2007}).

\bibitem[{\citenamefont{Suenram and Lovas}(1980)}]{Suenram:JACS102:7180}
\bibinfo{author}{\bibfnamefont{R.~D.} \bibnamefont{Suenram}} \bibnamefont{and}
  \bibinfo{author}{\bibfnamefont{F.~J.} \bibnamefont{Lovas}},
  \bibinfo{journal}{J. Am. Chem. Soc.} \textbf{\bibinfo{volume}{102}},
  \bibinfo{pages}{7180} (\bibinfo{year}{1980}).

\bibitem[{\citenamefont{Rizzo et~al.}(1985)\citenamefont{Rizzo, Park, Peteanu,
  and Levy}}]{Rizzo:JCP83:4819}
\bibinfo{author}{\bibfnamefont{T.~R.} \bibnamefont{Rizzo}},
  \bibinfo{author}{\bibfnamefont{Y.~D.} \bibnamefont{Park}},
  \bibinfo{author}{\bibfnamefont{L.}~\bibnamefont{Peteanu}}, \bibnamefont{and}
  \bibinfo{author}{\bibfnamefont{D.~H.} \bibnamefont{Levy}},
  \bibinfo{journal}{J. Chem. Phys.} \textbf{\bibinfo{volume}{83}},
  \bibinfo{pages}{4819} (\bibinfo{year}{1985}).

\bibitem[{\citenamefont{Snoek et~al.}(2001)\citenamefont{Snoek, Kroemer,
  Hockridge, and Simons}}]{Snoek:PCCP3:1819}
\bibinfo{author}{\bibfnamefont{L.~C.} \bibnamefont{Snoek}},
  \bibinfo{author}{\bibfnamefont{R.~T.} \bibnamefont{Kroemer}},
  \bibinfo{author}{\bibfnamefont{M.~R.} \bibnamefont{Hockridge}},
  \bibnamefont{and} \bibinfo{author}{\bibfnamefont{J.~P.}
  \bibnamefont{Simons}}, \bibinfo{journal}{Phys. Chem. Chem. Phys.}
  \textbf{\bibinfo{volume}{3}}, \bibinfo{pages}{1819} (\bibinfo{year}{2001}).

\bibitem[{\citenamefont{Hedberg et~al.}(1991)\citenamefont{Hedberg, Hedberg,
  Bethune, Brown, Dorn, Johnson, and de~Vries}}]{Hedberg:Science254:410}
\bibinfo{author}{\bibfnamefont{K.}~\bibnamefont{Hedberg}},
  \bibinfo{author}{\bibfnamefont{L.}~\bibnamefont{Hedberg}},
  \bibinfo{author}{\bibfnamefont{D.~S.} \bibnamefont{Bethune}},
  \bibinfo{author}{\bibfnamefont{C.~A.} \bibnamefont{Brown}},
  \bibinfo{author}{\bibfnamefont{H.~C.} \bibnamefont{Dorn}},
  \bibinfo{author}{\bibfnamefont{R.~D.} \bibnamefont{Johnson}},
  \bibnamefont{and} \bibinfo{author}{\bibfnamefont{M.}~\bibnamefont{de~Vries}},
  \bibinfo{journal}{Science} \textbf{\bibinfo{volume}{254}},
  \bibinfo{pages}{410} (\bibinfo{year}{1991}).

\bibitem[{\citenamefont{Chapman et~al.}(2006)\citenamefont{Chapman, Barty,
  Bogan, Boutet, Frank, Hau-Riege, Marchesini, Woods, Bajt, Benner, A.,
  Pl\"onjes, Kuhlmann, Treusch, D\"usterer, Tschentscher, Schneider, Spiller,
  M\"oller, Bostedt, Hoener, Shapiro, Hodgson, van~der Spoel, Burmeister,
  Bergh, Caleman, Huldt, Seibert, Maia, Lee, Sz\"onke, Timneanu, and
  Hajdu}}]{Chapman:NatPhys2:839}
\bibinfo{author}{\bibfnamefont{H.~N.} \bibnamefont{Chapman}},
  \bibinfo{author}{\bibfnamefont{A.}~\bibnamefont{Barty}},
  \bibinfo{author}{\bibfnamefont{M.~J.} \bibnamefont{Bogan}},
  \bibinfo{author}{\bibfnamefont{S.}~\bibnamefont{Boutet}},
  \bibinfo{author}{\bibfnamefont{S.}~\bibnamefont{Frank}},
  \bibinfo{author}{\bibfnamefont{S.~P.} \bibnamefont{Hau-Riege}},
  \bibinfo{author}{\bibfnamefont{S.}~\bibnamefont{Marchesini}},
  \bibinfo{author}{\bibfnamefont{B.~W.} \bibnamefont{Woods}},
  \bibinfo{author}{\bibfnamefont{S.}~\bibnamefont{Bajt}},
  \bibinfo{author}{\bibfnamefont{W.~H.} \bibnamefont{Benner}},
  \bibinfo{author}{\bibfnamefont{L.~W.} \bibnamefont{A.}},
  \bibinfo{author}{\bibfnamefont{E.}~\bibnamefont{Pl\"onjes}},
  \bibinfo{author}{\bibfnamefont{M.}~\bibnamefont{Kuhlmann}},
  \bibinfo{author}{\bibfnamefont{R.}~\bibnamefont{Treusch}},
  \bibinfo{author}{\bibfnamefont{S.}~\bibnamefont{D\"usterer}},
  \bibinfo{author}{\bibfnamefont{T.}~\bibnamefont{Tschentscher}},
  \bibinfo{author}{\bibfnamefont{J.~R.} \bibnamefont{Schneider}},
  \bibinfo{author}{\bibfnamefont{E.}~\bibnamefont{Spiller}},
  \bibinfo{author}{\bibfnamefont{T.}~\bibnamefont{M\"oller}},
  \bibinfo{author}{\bibfnamefont{C.}~\bibnamefont{Bostedt}},
  \bibinfo{author}{\bibfnamefont{M.}~\bibnamefont{Hoener}},
  \bibinfo{author}{\bibfnamefont{D.~A.} \bibnamefont{Shapiro}},
  \bibinfo{author}{\bibfnamefont{K.~O.} \bibnamefont{Hodgson}},
  \bibinfo{author}{\bibfnamefont{D.}~\bibnamefont{van~der Spoel}},
  \bibinfo{author}{\bibfnamefont{F.}~\bibnamefont{Burmeister}},
  \bibinfo{author}{\bibfnamefont{M.}~\bibnamefont{Bergh}},
  \bibinfo{author}{\bibfnamefont{C.}~\bibnamefont{Caleman}},
  \bibinfo{author}{\bibfnamefont{G.}~\bibnamefont{Huldt}},
  \bibinfo{author}{\bibfnamefont{M.~M.} \bibnamefont{Seibert}},
  \bibinfo{author}{\bibfnamefont{F.~R. N.~C.} \bibnamefont{Maia}},
  \bibinfo{author}{\bibfnamefont{R.~W.} \bibnamefont{Lee}},
  \bibinfo{author}{\bibfnamefont{A.}~\bibnamefont{Sz\"onke}},
  \bibinfo{author}{\bibfnamefont{N.}~\bibnamefont{Timneanu}}, \bibnamefont{and}
  \bibinfo{author}{\bibfnamefont{J.}~\bibnamefont{Hajdu}},
  \bibinfo{journal}{Nature Phys.} \textbf{\bibinfo{volume}{2}},
  \bibinfo{pages}{839} (\bibinfo{year}{2006}).

\bibitem[{\citenamefont{Gerlich et~al.}(2007)\citenamefont{Gerlich,
  Hackerm\"uller, Hornberger, Stibor, Ulbricht, Gring, Goldfarb, Savas, M\"uri,
  Mayor, and Arndt}}]{Gerlich:NatPhys3:711}
\bibinfo{author}{\bibfnamefont{S.}~\bibnamefont{Gerlich}},
  \bibinfo{author}{\bibfnamefont{L.}~\bibnamefont{Hackerm\"uller}},
  \bibinfo{author}{\bibfnamefont{K.}~\bibnamefont{Hornberger}},
  \bibinfo{author}{\bibfnamefont{A.}~\bibnamefont{Stibor}},
  \bibinfo{author}{\bibfnamefont{H.}~\bibnamefont{Ulbricht}},
  \bibinfo{author}{\bibfnamefont{M.}~\bibnamefont{Gring}},
  \bibinfo{author}{\bibfnamefont{F.}~\bibnamefont{Goldfarb}},
  \bibinfo{author}{\bibfnamefont{T.}~\bibnamefont{Savas}},
  \bibinfo{author}{\bibfnamefont{M.}~\bibnamefont{M\"uri}},
  \bibinfo{author}{\bibfnamefont{M.}~\bibnamefont{Mayor}}, \bibnamefont{and}
  \bibinfo{author}{\bibfnamefont{M.}~\bibnamefont{Arndt}},
  \bibinfo{journal}{Nature Phys.} \textbf{\bibinfo{volume}{3}},
  \bibinfo{pages}{711} (\bibinfo{year}{2007}).

\bibitem[{\citenamefont{Gordy and Cook}(1984)}]{Gordy:MWMolSpec}
\bibinfo{author}{\bibfnamefont{W.}~\bibnamefont{Gordy}} \bibnamefont{and}
  \bibinfo{author}{\bibfnamefont{R.~L.} \bibnamefont{Cook}},
  \emph{\bibinfo{title}{Microwave Molecular Spectra}} (\bibinfo{publisher}{John
  Wiley \& Sons}, \bibinfo{address}{New York, NY, USA}, \bibinfo{year}{1984}),
  \bibinfo{edition}{3rd} ed.

\bibitem[{\citenamefont{Vliegen et~al.}(2004)\citenamefont{Vliegen, W\"orner,
  Softley, and Merkt}}]{Vliegen:PRL92:033005}
\bibinfo{author}{\bibfnamefont{E.}~\bibnamefont{Vliegen}},
  \bibinfo{author}{\bibfnamefont{H.~J.} \bibnamefont{W\"orner}},
  \bibinfo{author}{\bibfnamefont{T.~P.} \bibnamefont{Softley}},
  \bibnamefont{and} \bibinfo{author}{\bibfnamefont{F.}~\bibnamefont{Merkt}},
  \bibinfo{journal}{Phys. Rev. Lett.} \textbf{\bibinfo{volume}{92}},
  \bibinfo{pages}{033005} (\bibinfo{year}{2004}).

\bibitem[{\citenamefont{Abd El~Rahim et~al.}(2005)\citenamefont{Abd El~Rahim,
  Antoine, Broyer, Rayane, and Dugourd}}]{AbdElRahim:JPCA109:8507}
\bibinfo{author}{\bibfnamefont{M.}~\bibnamefont{Abd El~Rahim}},
  \bibinfo{author}{\bibfnamefont{R.}~\bibnamefont{Antoine}},
  \bibinfo{author}{\bibfnamefont{M.}~\bibnamefont{Broyer}},
  \bibinfo{author}{\bibfnamefont{D.}~\bibnamefont{Rayane}}, \bibnamefont{and}
  \bibinfo{author}{\bibfnamefont{P.}~\bibnamefont{Dugourd}},
  \bibinfo{journal}{J. Phys. Chem. A} \textbf{\bibinfo{volume}{109}},
  \bibinfo{pages}{8507} (\bibinfo{year}{2005}).

\bibitem[{\citenamefont{K\"upper et~al.}(2006)\citenamefont{K\"upper, Haak,
  Wohlfart, and Meijer}}]{Kuepper:RSI77:016106}
\bibinfo{author}{\bibfnamefont{J.}~\bibnamefont{K\"upper}},
  \bibinfo{author}{\bibfnamefont{H.}~\bibnamefont{Haak}},
  \bibinfo{author}{\bibfnamefont{K.}~\bibnamefont{Wohlfart}}, \bibnamefont{and}
  \bibinfo{author}{\bibfnamefont{G.}~\bibnamefont{Meijer}},
  \bibinfo{journal}{Rev. Sci. Instrum.} \textbf{\bibinfo{volume}{77}},
  \bibinfo{pages}{016106} (\bibinfo{year}{2006}).

\bibitem[{\citenamefont{Western}()}]{Western:pgopher}
\bibinfo{author}{\bibfnamefont{C.~M.} \bibnamefont{Western}},
  \emph{\bibinfo{title}{Pgopher, a program for simulating rotational
  structure}}, \bibinfo{note}{{U}niversity of {B}ristol, {B}ristol, {UK}}.

\bibitem[{\citenamefont{Borst et~al.}(2001)\citenamefont{Borst, Korter, and
  Pratt}}]{Borst:CPL350:485}
\bibinfo{author}{\bibfnamefont{D.~R.} \bibnamefont{Borst}},
  \bibinfo{author}{\bibfnamefont{T.~M.} \bibnamefont{Korter}},
  \bibnamefont{and} \bibinfo{author}{\bibfnamefont{D.~W.} \bibnamefont{Pratt}},
  \bibinfo{journal}{Chem. Phys. Lett.} \textbf{\bibinfo{volume}{350}},
  \bibinfo{pages}{485} (\bibinfo{year}{2001}).

\bibitem[{\citenamefont{Wohlfart et~al.}(2008)\citenamefont{Wohlfart, Schnell,
  Grabow, and K\"upper}}]{Wohlfart:JMolSpec247:119}
\bibinfo{author}{\bibfnamefont{K.}~\bibnamefont{Wohlfart}},
  \bibinfo{author}{\bibfnamefont{M.}~\bibnamefont{Schnell}},
  \bibinfo{author}{\bibfnamefont{J.-U.} \bibnamefont{Grabow}},
  \bibnamefont{and} \bibinfo{author}{\bibfnamefont{J.}~\bibnamefont{K\"upper}},
  \bibinfo{journal}{J. Mol. Spec.} \textbf{\bibinfo{volume}{247}},
  \bibinfo{pages}{119} (\bibinfo{year}{2008}).

\bibitem[{\citenamefont{van Veldhoven et~al.}(2005)\citenamefont{van Veldhoven,
  Bethlem, and Meijer}}]{Veldhoven:PRL94:083001}
\bibinfo{author}{\bibfnamefont{J.}~\bibnamefont{van Veldhoven}},
  \bibinfo{author}{\bibfnamefont{H.~L.} \bibnamefont{Bethlem}},
  \bibnamefont{and} \bibinfo{author}{\bibfnamefont{G.}~\bibnamefont{Meijer}},
  \bibinfo{journal}{Phys. Rev. Lett.} \textbf{\bibinfo{volume}{94}},
  \bibinfo{pages}{083001} (\bibinfo{year}{2005}).

\end{thebibliography}

\end{document}